\documentclass[conference]{IEEEtran}
\IEEEoverridecommandlockouts
\usepackage{cite}
\usepackage{amsmath,amssymb,amsfonts}
\usepackage{algorithmic}
\usepackage{graphicx}
\usepackage{textcomp}
\usepackage{xcolor}
\usepackage{caption} 
\usepackage{booktabs}
\usepackage{multirow}
\usepackage{tcolorbox}
\usepackage{arydshln}
\usepackage{newfloat}
\usepackage{longtable}
\usepackage{listings}
\usepackage{rotating}
\usepackage{url}
\def\BibTeX{{\rm B\kern-.05em{\sc i\kern-.025em b}\kern-.08em
    T\kern-.1667em\lower.7ex\hbox{E}\kern-.125emX}}
    \makeatletter
\newcommand{\linebreakand}{
  \end{@IEEEauthorhalign}
  \hfill\mbox{}\par
  \mbox{}\hfill\begin{@IEEEauthorhalign}
}
\makeatother
\begin{document}

\title{Modeling the Attack: Detecting AI-Generated Text by Quantifying Adversarial Perturbations\\
}

\author{
    \IEEEauthorblockN{L. D. M. S. Sai Teja}
    \IEEEauthorblockA{\textit{Computer Science \& Engineering} \\
    \textit{National Institute of Technology}\\
    Silchar, India \\
    lekkalad\_ug\_22@cse.nits.ac.in}
    \and
    \IEEEauthorblockN{Annepaka Yadagiri}
    \IEEEauthorblockA{\textit{Computer Science \& Engineering} \\
    \textit{National Institute of Technology}\\
    Silchar, India \\
    annepaka22\_rs@cse.nits.ac.in}
    \and
    \IEEEauthorblockN{Sangam Sai Anish}
    \IEEEauthorblockA{\textit{Computer Science \& Engineering} \\
    \textit{National Institute of Technology}\\
    Silchar, India \\
    sangams\_ug\_22@cse.nits.ac.in}
    \linebreakand
    \IEEEauthorblockN{Siva Gopala Krishna Nuthakki}
    \IEEEauthorblockA{\textit{Computer Science \& Engineering} \\
    \textit{BML Munjal University}\\
    Haryana, India \\
    sivagopalkrishna04@gmail.com}
    \and
    \IEEEauthorblockN{Partha Pakray}
    \IEEEauthorblockA{\textit{Computer Science \& Engineering} \\
    \textit{National Institute of Technology}\\
    Silchar, India \\
    partha@cse.nits.ac.in}
}

\maketitle

\begin{abstract}
The growth of highly advanced Large Language Models (LLMs) constitutes a huge dual-use problem, making it necessary to create dependable AI-generated text detection systems. Modern detectors are notoriously vulnerable to adversarial attacks, with paraphrasing standing out as an effective evasion technique that foils statistical detection. This paper presents a comparative study of adversarial robustness, first by quantifying the limitations of standard adversarial training and then by introducing a novel, significantly more resilient detection framework: \textbf{Perturbation-Invariant Feature Engineering (PIFE)}, a framework that enhances detection by first transforming input text into a standardized form using a multi-stage normalization pipeline, it then quantifies the transformation's magnitude using metrics like Levenshtein distance and semantic similarity, feeding these signals directly to the classifier. We evaluate both a conventionally hardened Transformer and our PIFE-augmented model against a hierarchical taxonomy of character-, word-, and sentence-level attacks. Our findings first confirm that conventional adversarial training, while resilient to syntactic noise, fails against semantic attacks, an effect we term ``\textit{semantic evasion threshold}'', where its True Positive Rate at a strict 1\% False Positive Rate plummets to 48.8\%. In stark contrast, our PIFE model, which explicitly engineers features from the discrepancy between a text and its canonical form, overcomes this limitation. It maintains a remarkable \textbf{82.6\% TPR} under the same conditions, effectively neutralizing the most sophisticated semantic attacks. This superior performance demonstrates that explicitly modeling perturbation artifacts, rather than merely training on them, is a more promising path toward achieving genuine robustness in the adversarial arms race.
\end{abstract}


\begin{figure*}[h]
    \centering
    \resizebox{1.0\linewidth}{!}{%
    \includegraphics[width=1.0\textwidth]{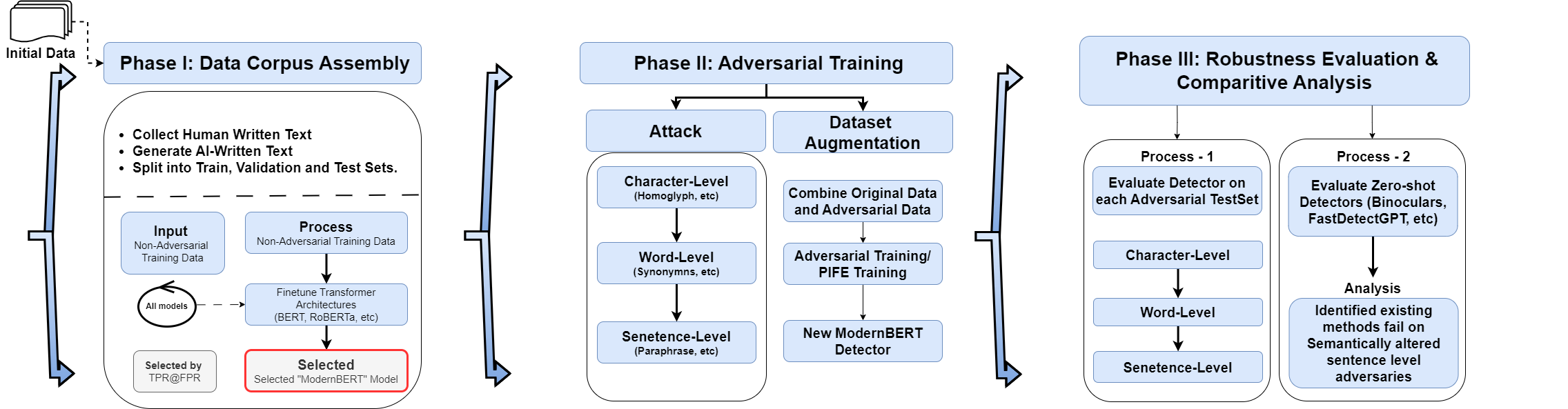}
    }
    \caption{Workflow of the Entire pipeline process for both modeling and evaluation.}
    \label{fig:pipeline}
\end{figure*}

\section{Introduction}
The sudden emergence of Large Language Models (LLMs) is a paradigm shift for the Natural Language Processing (NLP) community \cite{kasneci2023chatgpt}. The models showcase an exceptional ability to produce text that not only is smooth and coherent but is also frequently indistinguishable from human-written texts across a broad range of applications, ranging from literary content to technical manuals. This innovation, while awesome, poses a daunting dual-use problem. As much as LLMs pose unprecedented opportunities for creativity and productivity, they also pose significant societal threats. Malicious use, such as automatic creation of propaganda and disinformation, copyright infringement, the ability to create sophisticated phishing campaigns, and the erosion of academic honesty, requires the development of strong means of AI-generated text detection.

The terrain of AI text detection today is fraught with complications. Many studies and real-world implementations have shown that current detection software, both commercial and open-source, is usually not reliable. They may fail to detect content from state-of-the-art (SOTA) LLMs and often have high levels of biases, for example, a greater tendency to mislabel text composed by non-native English speakers as AI-written. This lack of reliability is a main hindrance to their general use in high-stakes settings where the penalty for a misclassification can be dramatic. Adding to this challenge is the growing threat of adversarial attacks. These are not fortuitous errors but intentional, well-designed alterations to AI-written text specifically aimed at bypassing detection systems. Of the many evasion methods, paraphrasing has proven to be a particularly strong threat. By paraphrasing an AI-produced message, an attacker can dramatically change its statistical attributes while maintaining its fundamental semantic meaning, thus drastically lowering the accuracy of numerous detection tools. This phenomenon generates an adversarial `arms race,' as improvements in detection technology are constantly matched with increasingly clever evasion techniques. Therefore, the issue is not so much to construct a correct classifier but to design a robust system that can resist attacks from a clever and adaptive opponent.

This paper confronts these challenges through a systematic study of the adversarial robustness of a supervised, Transformer-based detector. The main contributions of this work are fourfold:

\begin{enumerate}
    \item \textit{We conduct a rigorous baseline evaluation of prominent Transformer architectures to identify an optimal base model for the AI text detection task.}

    \item \textit{We introduce a novel defense framework, \textbf{Perturbation-Invariant Feature Engineering (PIFE)}, which explicitly models adversarial artifacts by computing a discrepancy vector between an input text and its canonical, normalized form.}

    \item \textit{We present a comprehensive comparative analysis, evaluating both a standard adversarially trained detector and our PIFE-augmented model against a hierarchical taxonomy of character-, word-, and sentence-level attacks to quantify their respective robustness.}

    \item \textit{We empirically demonstrate that while conventional adversarial training fails against sophisticated semantic attacks, our PIFE model successfully overcomes the established “semantic evasion threshold,” achieving state-of-the-art robustness, particularly against paraphrasing.}
\end{enumerate}

\section{Problem Formulation and Research Scope}

\subsection{Binary Classification of AI-Generated Text}  
The fundamental task addressed in this research is the binary classification of a given text's origin. Formally, given a text sequence \( X \) composed of tokens \( (x_1, x_2, \ldots, x_n) \), the objective is to learn a classification function \( f(X) \) that maps the input sequence to one of two discrete labels: $f(X) \in \{ \text{Human}, \text{AI} \}$. This formulation permits the use of common supervised learning methods and metrics of performance. To offer a complete picture of model performance, this work utilizes a battery of evaluation metrics. Common classification metrics such as overall Accuracy, class-wise Precision, Recall, and F1-Score are utilized to offer a detailed view of the model's performance on human- as well as AI-generated text. The Area Under the Receiver Operating Characteristic Curve (AUROC) is used to evaluate the overall discriminative power of the model across all potential classification thresholds. But in actual uses of AI text detection, such as maintaining academic honesty or detecting state-sponsored disinformation campaigns, the social and individual cost of a false positive (mistakenly labeling human-written text as AI-generated) tends to be much greater than that of a false negative.

A learner mistakenly accused of plagiarism by an error of a detector is severely penalized, so a low False Positive Rate (\textit{FPR}) is not an option for ethical deployment. It follows that measures such as \textit{AUROC}, which average performance over the whole range of \textit{FPRs}, can be deceptive, since a high value can cover up for bad performance at the particular low-\textit{FPR} operating points necessitated by practical application. To counter this, the True Positive Rate (\textit{TPR}) at fixed, low \textit{FPR} thresholds (in particular 5\%, 3\%, and 1\%) is used as a main measure. \textit{TPR@FPR} measures the effectiveness of a detector in terms of a high standard of evidence with the question: ``\textit{At an acceptably low false accusation rate, what percentage of true AI-generated text can the system correctly identify?}'' This measure gives a more practical and responsible measure of the real-world effectiveness of a detector. In addition to nominal classification accuracy, the focus of this work is primarily on adversarial robustness: the robustness of a detector to remain accurate in the face of inputs that have been deliberately altered to induce misclassification.

\subsection{Adversarial Robustness} 
\label{subsec:adv_robust}
An adversarial attack in the NLP context is applying fine, usually meaning-preserving, perturbations to a text sequence in an effort to go undetected. The attacks may be categorized according to what the adversary knows about the target model, going from white-box situations where the attacker possesses complete access to the model's architecture and parameters to black-box situations where the attacker only has access to querying the model and seeing its outputs. In order to critically test the worst-case resilience of our model, in this research, a white-box adversarial example generation method is used, thus stress-testing the detector against an adversarially maximally informed opponent. 

To systematically analyze the model's vulnerabilities, the adversarial attacks are organized into a three-level hierarchy based on the scope and complexity of the textual perturbations applied. This taxonomy allows for a structured investigation into how different types of manipulations affect the detector's performance.
\begin{enumerate}
    \item \textbf{\textit{Character-Level Attacks:}} These involve minor modifications at the sub-word level that are often imperceptible to human readers but can disrupt the model's tokenization and input processing. Examples investigated in this study include character deletions, insertions, and swaps; homoglyph attacks, which replace characters with visually similar Unicode characters; the insertion of invisible characters; and simulated keyboard typos.
    \item \textbf{\textit{Word-Level Attacks:}} These perturbations operate at the word level, targeting the syntactic structure and local semantic content of the text. Examples include the replacement of words with their synonyms or antonyms, random word deletion and insertion, and the reordering of words within a sentence.
    \item \textbf{\textit{Sentence-Level Attacks:}} This category comprises the most sophisticated attacks, which involve global, semantic-preserving transformations that fundamentally alter the text's phrasing and structure while retaining its original meaning. These attacks are widely recognized in the literature as being particularly effective at evading detection. The attacks evaluated include paraphrasing, which rewrites sentences or entire passages; tense alteration; and the reordering, splitting, or fusion of sentences.
\end{enumerate}

\section{Related Work}
\label{sec:rel_work}
Adversarial robustness in AI-generated text (AIGT) detection has attracted increasing attention as detectors face evasion through paraphrasing, syntactic modification, or embedding-level perturbations. A series of works demonstrates that paraphrasing or humanizer-style rewrites can drastically undermine detection accuracy. \cite{b3} shows that paraphrasing significantly reduces the reliability of leading detectors, and proposes retrieval-based defenses, while \cite{b4} constructs adversarial rewriting pipelines to evade detection. \cite{b5} further introduces a universal adversarial attack that humanizes machine outputs across multiple generators, highlighting the need for more robust detection. To mitigate such attacks, robust detector architectures have been explored. \cite{b6} propose the Siamese Calibrated Reconstruction Network (SCRN), which learns perturbation-invariant representations and resists character- and word-level noise. \cite{b7} improves upon DetectGPT by introducing Fast-DetectGPT, which accelerates zero-shot detection using conditional probability curvature. These approaches demonstrate how reconstruction-based and probabilistic curvature-based techniques can enhance robustness beyond surface-level statistics. Another line of research focuses on embedding- and token-probability-based adversarial attacks. \cite{b8} design embedding-level perturbations that manipulate token probability signals to deceive detectors. Complementary studies propose adversarial frameworks to evaluate detector vulnerabilities under both black-box and white-box settings \cite{b9}. Together, these works emphasize that robustness must be considered not only against natural paraphrasers but also against fine-grained adversarial manipulations at the representation level. Zero-shot detectors have also been influential in the robustness discussion. \cite{b10} introduces DetectGPT, which relies on probability curvature around text sequences to distinguish human and machine writing. \cite{b11} proposes GLTR, offering statistical and visualization-based cues for detection. More recent evaluations reveal that zero-shot detectors suffer from sensitivity to domain and generator shifts, as studied in \cite{b12} and \cite{b13}. Interestingly, \cite{b14} shows that smaller language models can act as stronger black-box detectors, challenging assumptions about model size and robustness. Finally, data-centric and active-learning approaches provide a complementary defensive strategy. \cite{b15} introduces the DAMAGE framework, which augments training data with syntactically humanized adversarial examples and leverages active learning for improved generalization. Retrieval-based defenses \cite{b3} similarly strengthen detectors against paraphrasing, while broader surveys \cite{b16} summarize adversarial attacks and defense strategies in NLP, contextualizing robustness challenges faced by AIGT detection. Despite these advances, most prior work focuses on pure AI-generated text, with limited exploration of mixed-text segmentation where adversarial perturbations interact with human-authored passages.

\section{Dataset Description}\label{sec:data_desc}
The dataset employed in this study is sourced from the \textbf{\textit{CLEF 2024 PAN-Generative AI Authorship}} \cite{bevendorff2024overview} shared task on fake news detection. It is composed of texts based on U.S. news headlines from 2021 and is divided into two primary categories: human-written and AI-generated. The collection contains 1,087 text samples authored by humans and a total of 14,131 samples generated by 13 distinct LLMs. Each of the 13 models, including \texttt{alpaca-7b, bigscience-bloomz-7b1, chav-inlo-alpaca-13b, gemini-pro, gpt-3.5-turbo-0125, gpt-4-turbo-preview, metallama-2-7b, metallama270b, mistralai-mistral-7b-instruct, mistralai-mixtral-8x7b, qwem-qwen1.5-72b, text-bison-002, vicgalle-gpt2-open-instruct-v1,} contributed 1,087 text samples. For the purpose of training and evaluation, the entire dataset was first shuffled and undergoes a stratified splitting strategy, allocating 70\% of the data for training, 20\% for validation, and the remaining 10\% for testing. This approach ensures that the proportional representation of human-written text and text from each of the 13 LLMs is consistently maintained across all three subsets. Furthermore, to evaluate robustness, the entire text corpus was subjected to the adversarial attacks detailed in Table~\ref{tab:attack_summary}.

\begin{table}[ht]
\centering
\caption{Summary of Adversarial Attack Types and Methods}
\label{tab:attack_summary}
\resizebox{1.0\linewidth}{!}{%
\begin{tabular}{ll}
\toprule
\textbf{Attack Type} & \textbf{Attacks} \\
\midrule
\multirow{3}{*}{\textbf{Character-Level}} & Char Deletion, Char Insertion, Char Swap, \\
 & Homoglyph, Invisible Char, Keyboard Typo, \\
 & Punctuation, Upper-Lower, All Mix \\
\midrule
\multirow{2}{*}{\textbf{Word-Level}} & Synonym Replacement, Antonym Replacement, Word Deletion \\
 & Word Insertion, Word Reordering, All Mix \\
\midrule
\multirow{2}{*}{\textbf{Sentence-Level}} & Paraphrase, Tense Altering, Sentence Reordering \\
 & Sentence Splitting, Sentence Fusion, All Mix \\
\bottomrule
\end{tabular}
}
\end{table}

\section{Methods}
\label{sec:method}

\subsection{Adversarial Training of a Supervised Detector}
The experimental methodology of this study is centered on the fine-tuning and adversarial training of a supervised, Transformer-based classifier. The workflow of the pipeline for the modeling and evaluation is shown in Figure~\ref{fig:pipeline}.

\begin{figure}[h]
    \centering
    \includegraphics[width=0.8\linewidth]{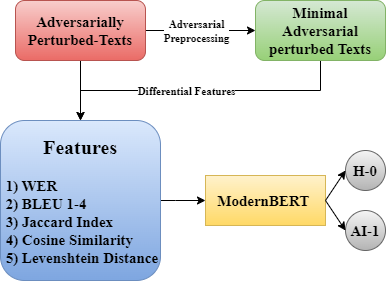}
    \caption{Workflow of Perturbation-Invariant Feature Engineering}
    \label{fig:pife}
\end{figure}

\begin{enumerate}
    \item \textbf{\textit{Baseline Model Architectures:}} The initial phase of the research involved a comparative evaluation of several prominent Transformer-based architectures to establish a performance baseline. This included models from the \textit{BERT} family, which are based on a bidirectional Transformer encoder architecture that processes text in its full left and right context simultaneously. Also included were models from the \textit{RoBERTa} family, which share \textit{BERT's} architecture but benefit from a more robustly optimized pretraining procedure. The models included in this phase are BERT \cite{devlin2019bert}, RoBERTa \cite{liu2019roberta}, DistilBERT \cite{sanh2019distilbert}, XLNET \cite{yang2019xlnet}, ALBERT \cite{lan2019albert}, DeBERTa \cite{he2021debertav3}, and ModernBERT \cite{warner2024smarter}, their number of parameters and HuggingFace sources are given in the Table~\ref{tab:model_params}.
    
    \item \textbf{\textit{Adversarial Training Protocol:}} To enhance the detector's resilience against evasion attempts, an adversarial training protocol was implemented. This technique functions as a form of targeted data augmentation. First, a large corpus of adversarial examples was generated by applying the full suite of character-, word-, and sentence-level attacks described in Section \ref{subsec:adv_robust} to a set of AI-generated texts. The standard training dataset was then augmented with these adversarial examples. The \textit{ModernBERT} classifier was subsequently fine-tuned on this expanded dataset, which contained original human texts, original AI texts, and adversarially perturbed AI texts. This process explicitly exposes the model to the patterns and artifacts introduced by adversarial attacks, compelling it to learn representations that are invariant to such perturbations. This approach can be conceptualized as a `\textit{vaccination}' for the model, preemptively teaching it to recognize and correctly classify malicious inputs, thereby hardening it against future attacks. 

    \item \textbf{\textit{Perturbation-Invariant Feature Engineering (PIFE):}} 
    As a novel alternative to the data augmentation approach of adversarial training, we introduce a feature engineering methodology designed to explicitly model and quantify the artifacts introduced by adversarial attacks. This technique, termed Perturbation-Invariant Feature Engineering (\textbf{\textit{PIFE}}), operates on the hypothesis that adversarial perturbations create a measurable discrepancy between a manipulated text and its canonical, preprocessed form. The pipeline is architected as shown in Figure~\ref{fig:pife} and the pipeline goes as follows: 
    \begin{enumerate}
        \item Text Canonicalization: Given an input text sequence $X$, which may be either pristine or adversarially perturbed, we first apply a normalization function, $N(.)$. This function is designed to neutralize common adversarial manipulations and produce a canonical version of the text, $X' =N(X)$,
        \item Discrepancy Vector Computation: We then engineer a discrepancy vector, $\mathbf{v}_d$, by computing a suite of comparative metrics between the original text $X$ and its canonical counterpart $X'$. This vector serves to quantify the magnitude and nature of the perturbation. The features comprising $\mathbf{v}_d$ include: a) Cosine Similarity: Between the sentence embeddings of $X$ and $X'$, to quantify the degree of semantic shift introduced by the perturbation., b) Levenshtein Distance: To capture fine-grained, character- and word-level edits., c) Jaccard Index: To measure the overlap of vocabulary between the original and canonical texts, d) BLEU Score \& Word Error Rate (WER): To assess the structural and n-gram similarity, which is particularly sensitive to reordering attacks, 
        \item Augmented Input Representation: The classifier input combines the token embeddings of text $X$ with the discrepancy vector $\mathbf{v}_d$, providing both semantic content and a quantitative signal of potential manipulation, 
        \item Implicit Adversarial Inference: Critically, The model isn’t given an explicit attack indicator; instead, it learns end-to-end to associate patterns in the discrepancy vector $\mathbf{v}_d$ with the origin label ($y \in \{\text{Human, AI}\}$), allowing it to adapt its classification based on perturbation strength.
    \end{enumerate}  
    
    As shown in Table~\ref{tab:pife_attack_performance}, this PIFE-augmented model significantly outperforms the standard adversarially trained architecture. The hyperparameters taken for this method are given in Table~\ref{tab:hyperparameters}. The actual workflow of \textit{PIFE} can be seen in the Figure~\ref{fig:pife}.
\end{enumerate}

\begin{table}[ht]
\centering
\caption{Soures of Transformer Model with Parameters}
\label{tab:model_params}
\resizebox{1.0\linewidth}{!}{%
\begin{tabular}{lcl}
\toprule
\textbf{Model} & \textbf{\# Params} & \textbf{Hugging Face Link} \\
\midrule
BERT         & 110M                   & \url{https://huggingface.co/google-bert/bert-base-uncased}      \\
DistilBERT   & 66M                    & \url{https://huggingface.co/distilbert/distilbert-base-uncased}   \\
RoBERTa      & 125M                   & \url{https://huggingface.co/FacebookAI/roberta-base}           \\
XLNET        & 117M                   & \url{https://huggingface.co/xlnet/xlnet-base-cased}           \\
ALBERT       & 12M                    & \url{https://huggingface.co/albert/albert-base-v2}            \\
DeBERTa      & 184M                   & \url{https://huggingface.co/microsoft/deberta-v3-base}   \\
ModernBERT   & 149M                   & \url{https://huggingface.co/answerdotai/ModernBERT-base}\\
\bottomrule
\end{tabular}
}
\end{table}

\subsection{Hyperparameters}
The model training and evaluation were conducted using a specific set of hyperparameters. For our base architecture, we employed a pre-trained, Transformer-based language model. For the input text, we set the maximum sequence length to \textit{512} tokens, with shorter texts being padded and longer texts truncated to ensure uniform input size. The optimization was performed using the \textit{AdamW} optimizer with a learning rate of $2 \times 10^{-5}$. The model was trained with a batch size of \textit{32}. For this binary classification task, we employed the \textit{CrossEntropyLoss} function as our loss criterion. The training was scheduled for a maximum of \textit{5} epochs. To prevent overfitting, we implemented an early stopping mechanism with a patience of \textit{2}. This strategy halts the training process if the validation loss does not show improvement for two consecutive epochs, and the model weights from the epoch with the lowest validation loss are restored for the final evaluation.

\begin{table}[h!]
\centering
\caption{Hyperparameters for Model Training}
\label{tab:hyperparameters}
\begin{tabular}{ll}
\toprule
\textbf{Hyperparameter}      & \textbf{Value}                 \\ 
\midrule
Base Model                   & Transformer-based Model        \\
Max Sequence Length          & 512                            \\
Optimizer                    & AdamW                           \\
Learning Rate                & $2 \times 10^{-5}$             \\
Batch Size                   & 32                             \\
Loss Function                & CrossEntropyLoss               \\
Epochs (Max)                 & 5                              \\
Early Stopping Patience      & 2                              \\ 
\bottomrule
\end{tabular}
\end{table}

\section{Comparisons}
\subsection{Open-Source Zero-Shot Detectors: A Review of the State-of-the-Art}
To provide a robust context for the performance of our supervised model, it is essential to review the current landscape of zero-shot AI text detectors. These methods are notable for their ability to detect AI-generated text without requiring task-specific training on a labeled dataset of human and AI examples. Instead, they leverage intrinsic statistical properties of text generated by LLMs.

\begin{enumerate}
    \item \textbf{\textit{FastDetectGPT}} \cite{bao2023fast}: This method is an efficient, curvature-based detector that builds upon its predecessor, DetectGPT. It operates on the hypothesis that text generated by an LLM tends to occupy regions of high positive curvature in the model's log-probability space. In other words, the log probability of a machine-generated token is typically higher than the average log probability of other plausible alternative tokens in that context. FastDetectGPT quantifies this curvature by efficiently sampling alternative tokens and comparing their log probabilities to that of the original text, providing a score that indicates the likelihood of machine generation.
    \item \textbf{\textit{Glimpse}} \cite{bao2023fast}: Glimpse addresses a key limitation of many powerful `\textit{white-box}' detection methods: their reliance on access to a model's full probability distributions, which is not available for proprietary, API-gated LLMs like GPT-4. Glimpse introduces a Probability Distribution Estimation (\textit{PDE}) technique that reconstructs an approximation of the full token probability distribution from the limited top-k probability outputs provided by these APIs. This enables the application of sophisticated white-box methods in a black-box setting, effectively bridging the gap between open-source and proprietary models for detection tasks. 
    \item \textbf{\textit{Binoculars}} \cite{hans2024spotting}: This is a novel zero-shot approach that requires no training data and instead utilizes a pair of pretrained LLMs to create a detection signal. The method, so-named because it views text through two `\textit{lenses,}' calculates a score based on the ratio of a text's perplexity as measured by an `\textit{observer}' LLM to its cross-perplexity, where the `\textit{performer}' LLM's predictions are evaluated by the observer. This contrastive measure has proven to be a highly accurate signal for distinguishing between the predictable, low-perplexity nature of machine text and the more varied, higher-perplexity nature of human writing. 
    \item \textbf{\textit{LogRank}} \cite{su2023detectllm}: This family of statistical methods leverages the rank of a token in a model's predicted vocabulary distribution, rather than its absolute probability. The core intuition is that LLMs tend to sample tokens that are consistently ranked high (for example, within the top-k choices), whereas human word choice is less constrained. By analyzing the distribution of token ranks (or log-ranks), these methods can identify statistical signatures of machine generation. Variants like Log-Likelihood Log-Rank Ratio (\textit{LRR}) combine rank information with probability information to create a more robust detection metric. 
\end{enumerate}

\section{Results and Discussions}
\label{sec:res_disc}
\begin{figure}[h]
    \centering
    \includegraphics[width=1.0\linewidth]{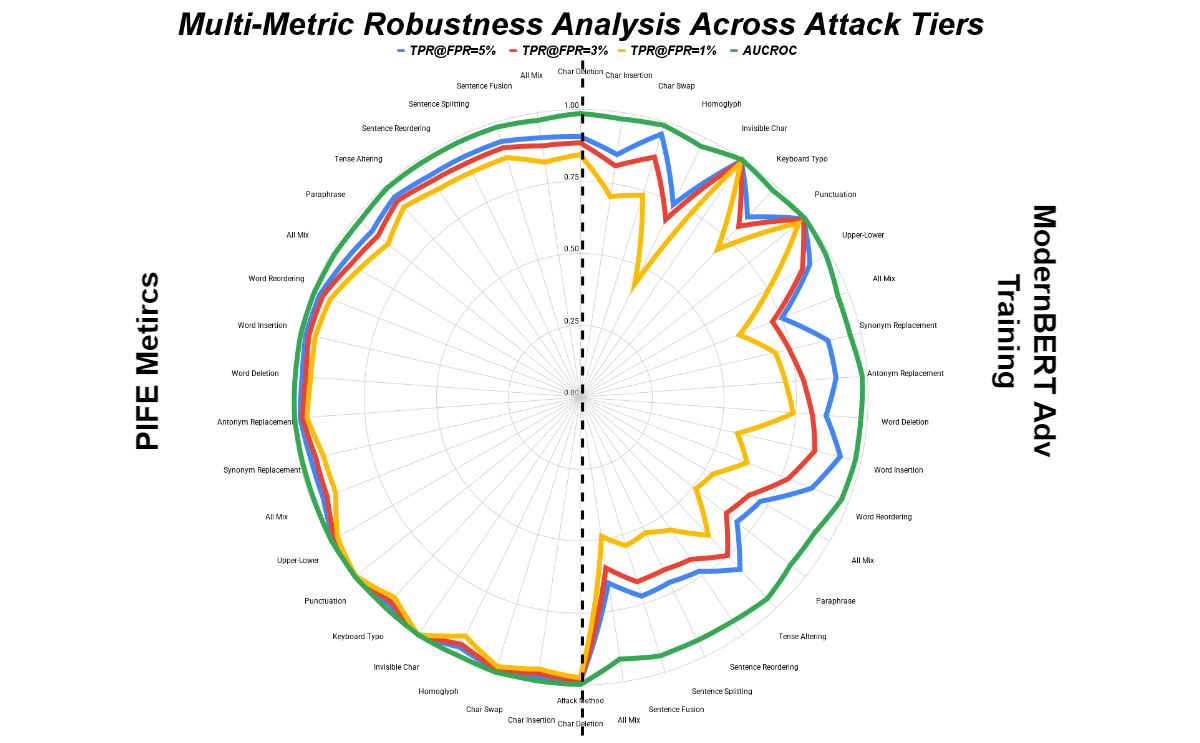}
    \caption{Radar chart metric visualization of both PIFE trained ModernBERT (left) and Adversarially Trained ModernBERT (right). The colored lines plot the model's performance scores against various text-based attacks, where results closer to the outer edge signify greater resilience.}
    \label{fig:metrics}
\end{figure}

The empirical evaluation of our proposed models was conducted in two primary stages. First, we performed a baseline comparison on a non-adversarial dataset to identify the most effective base architecture. Second, we conducted a comprehensive stress test comparing the robustness of a standard adversarially trained model against our novel Perturbation-Invariant Feature Engineering (PIFE) approach, using a hierarchical taxonomy of attacks. The results in the Radar plot can be seen in the Figure~\ref{fig:metrics}, where the left are the metrics in each attack by the PIFE model, and the right are the metrics in each attack with the Adversarially trained ModernBERT model.

\subsection{Baseline Performance on Non-Adversarial Data}
The initial experiments aimed to identify the most effective base architecture for the AI text detection task. As shown in Table \ref{tab:tpr_fpr_non_adv}, the `\textit{ModernBERT}' model demonstrated unequivocally superior performance. It achieved the highest \textit{AUROC} of 0.994 and, most impressively, a \textit{TPR} of 0.943 at a stringent \textit{FPR} of 1\%. A class-wise analysis in Table \ref{tab:non_adv_traditional} further reinforced its selection, as it achieved a more balanced F1-Score for both Human (0.897) and AI (0.992) classes. This strong, balanced performance justified its selection as the base architecture for subsequent adversarial robustness studies.

\begin{table}[htbp]
\centering
\caption{Baseline performance evaluation for binary classification on a non-adversarial dataset. We compare the True Positive Rate (\textit{TPR}) at fixed False Positive Rate (\textit{FPR}) thresholds of 5\%, 3\%, and 1\%, alongside the overall Area Under the ROC Curve (\textit{AUCROC}). The best performance for each metric is highlighted in bold.}
\label{tab:tpr_fpr_non_adv}
\resizebox{1.0\linewidth}{!}{%
\begin{tabular}{lcccc}
\toprule
\textbf{Model} & \textbf{\shortstack{TPR@FPR\\=5\%}} & \textbf{\shortstack{TPR@FPR\\=3\%}} & \textbf{\shortstack{TPR@FPR\\=1\%}} & \textbf{\shortstack{AUC\\ROC}} \\
\midrule
BERT       & 0.945          & 0.940          & 0.887          & 0.990           \\
DistilBERT & 0.788          & 0.761          & 0.743          & 0.973           \\
RoBERTa    & 0.958          & 0.951          & 0.873          & 0.993           \\
XLNET      & 0.955          & 0.946          & 0.887          & 0.992           \\
ALBERT     & 0.928          & 0.905          & 0.841          & 0.987           \\
DeBERTa    & 0.961          & 0.932          & 0.739          & 0.988           \\
ModernBERT & \textbf{0.973} & \textbf{0.955} & \textbf{0.943} & \textbf{0.994}  \\
\bottomrule
\end{tabular}
}
\end{table}

\begin{table}[htbp]
\centering
\caption{Class-wise performance using traditional binary classification metrics on a non-adversarial dataset. The best scores are highlighted in bold.}
\label{tab:non_adv_traditional}
\resizebox{1.0\linewidth}{!}{%
\begin{tabular}{lccccccc}
\toprule
& \multicolumn{3}{c}{\textbf{Human}} & \multicolumn{3}{c}{\textbf{AI}} & \\
\cmidrule(lr){2-4} \cmidrule(lr){5-7}
\textbf{Model} & \textbf{Precision} & \textbf{Recall} & \textbf{F1-Score} & \textbf{Precision} & \textbf{Recall} & \textbf{F1-Score} & \textbf{Accuracy} \\
\midrule
BERT       & 0.930 & 0.733 & 0.820 & 0.979 & 0.995 & 0.987 & 0.976 \\
DistilBERT & 0.907 & 0.449 & 0.601 & 0.959 & 0.996 & 0.977 & 0.957 \\
RoBERTa    & 0.943 & 0.761 & 0.842 & 0.981 & 0.996 & 0.989 & 0.979 \\
XLNET      & 0.914 & 0.688 & 0.810 & 0.976 & 0.993 & 0.987 & 0.976 \\
ALBERT     & 0.868 & 0.788 & 0.826 & 0.983 & 0.990 & 0.987 & 0.976\\
DeBERTa    & 0.951 & 0.715 & 0.816 & 0.978 & 0.997 & 0.987 & 0.976\\
ModernBERT & \textbf{0.986} & \textbf{0.880} & \textbf{0.897} & \textbf{0.990} & \textbf{0.999} & \textbf{0.992} & \textbf{0.985} \\
\bottomrule
\end{tabular}
}
\end{table}

\subsection{Comparative Analysis of Adversarial Robustness}
Following the baseline evaluation, we compared two defense strategies: the standard \textbf{Adversarial Training} protocol applied to \textit{ModernBERT}, and our novel \textbf{PIFE-augmented model}. The results, presented in Table \ref{tab:all_level_attacks} (Adversarial Training) and Table \ref{tab:pife_attack_performance} (PIFE), reveal a dramatic difference in robustness, particularly against sophisticated attacks.

\begin{enumerate}
    \item \textbf{Character-Level Robustness:} ModernBERT handled simple character attacks well (AUROC $\approx$ 0.998), but struggled with homoglyphs (0.967) and mixed attacks (TPR@FPR=1\%: 0.593). In contrast, PIFE was consistently strong, maintaining AUROC $\geq$ 0.992 and a much higher 0.912 TPR@FPR=1\% under All Mix.  
    \item \textbf{Word-Level Robustness:} ModernBERT weakened under word-level attacks like synonym replacement (AUROC 0.962) and especially All Mix (TPR@FPR=1\%: 0.533). PIFE remained robust, reaching AUROC 0.988 and TPR@FPR=1\%: 0.887, showing the benefit of explicitly modeling word perturbations.
    \item \textbf{Sentence-Level Robustness:} Sentence-level paraphrasing nearly broke ModernBERT (TPR@FPR=1\% $\approx$ 0.512, All Mix $\approx$ 0.488). PIFE, however, excelled by comparing texts to canonical forms, achieving AUROC 0.981 and TPR@FPR=1\%: 0.854. Even under All Mix, it sustained 0.826, proving far more resilient to semantic-preserving attacks.  
\end{enumerate}
ModernBERT resists simple noise but fails against semantic-level changes, revealing a ``semantic evasion threshold.'' PIFE overcomes this by directly modeling perturbations, making robustness explicit rather than implicit.

\begin{table}[ht]
\centering
\caption{Character-level adversarial attack performance for a \textit{ModernBERT} model fine-tuned on an augmented dataset of original and adversarial pairs. Results are reported on the held-out test set.}
\label{tab:all_level_attacks}
\resizebox{1.0\linewidth}{!}{%
\begin{tabular}{lllcccc}
\toprule
& \textbf{\shortstack{Attack\\Type}} & \textbf{\shortstack{Adv\\Attack}} & \textbf{\shortstack{TPR@FPR\\=5\%}} & \textbf{\shortstack{TPR@FPR\\=3\%}} & \textbf{\shortstack{TPR@FPR\\=1\%}} & \textbf{\shortstack{AUC\\ROC}} \\
\midrule
\multirow{21}{*}{\rotatebox{90}{\shortstack{Adversarially Trained ModernBERT Detector}}} 
 & \multirow{9}{*}{\rotatebox{90}{\shortstack{Character-Level}}} 
    & Char Deletion     & 0.905 & 0.884 & 0.843 & 0.986 \\
 &  & Char Insertion    & 0.853 & 0.812 & 0.705 & 0.978 \\
 &  & Char Swap         & 0.956 & 0.872 & 0.735 & 0.988 \\
 &  & Homoglyph         & 0.743 & 0.685 & 0.435 & 0.967 \\
 &  & Invisible Char    & 0.993 & 0.991 & 0.987 & 0.998 \\
 &  & Keyboard Typo     & 0.854 & 0.809 & 0.700 & 0.982 \\
 &  & Punctuation       & 0.993 & 0.992 & 0.969 & 0.998 \\
 &  & Upper-Lower       & 0.922 & 0.890 & 0.758 & 0.987 \\
 &  & All Mix           & 0.752 & 0.718 & 0.593 & 0.961 \\
\cmidrule{2-7}
 & \multirow{6}{*}{\rotatebox{90}{\shortstack{Word-Level}}}      
    & Synonym Replacement & 0.883 & 0.745 & 0.696 & 0.962 \\
 &  & Antonym Replacement & 0.891 & 0.777 & 0.715 & 0.983 \\
 &  & Word Deletion       & 0.856 & 0.809 & 0.742 & 0.977 \\
 &  & Word Insertion      & 0.926 & 0.836 & 0.561 & 0.981 \\
 &  & Word Reordering     & 0.863 & 0.775 & 0.621 & 0.975 \\
 &  & All Mix             & 0.723 & 0.679 & 0.533 & 0.940 \\
\cmidrule{2-7}
 & \multirow{6}{*}{\rotatebox{90}{\shortstack{Sentence-Level}}}  
    & Paraphrase          & 0.695 & 0.648 & 0.512 & 0.935 \\
 &  & Tense Altering      & 0.815 & 0.751 & 0.654 & 0.955 \\
 &  & Sentence Reordering & 0.733 & 0.682 & 0.559 & 0.941 \\
 &  & Sentence Splitting  & 0.715 & 0.667 & 0.523 & 0.938 \\
 &  & Sentence Fusion     & 0.724 & 0.671 & 0.540 & 0.940 \\
 &  & All Mix             & 0.654 & 0.601 & 0.488 & 0.921 \\
\bottomrule
\end{tabular}
}
\end{table}

\begin{table}[ht]
\centering
\caption{Adversarial attack performance for the \textbf{PIFE-augmented model}. The model demonstrates significant robustness improvements across all attack levels, particularly against semantic-preserving transformations. Results are reported on the held-out test set.}
\label{tab:pife_attack_performance}
\resizebox{1.0\linewidth}{!}{%
\begin{tabular}{lllcccc}
\toprule
& \textbf{\shortstack{Attack\\Type}} & \textbf{\shortstack{Adv\\Attack}} & \textbf{\shortstack{TPR@FPR\\=5\%}} & \textbf{\shortstack{TPR@FPR\\=3\%}} & \textbf{\shortstack{TPR@FPR\\=1\%}} & \textbf{\shortstack{AUC\\ROC}} \\
\midrule
\multirow{21}{*}{\rotatebox{90}{\shortstack{PIFE-Augmented with ModernBERT}}} 
 & \multirow{9}{*}{\rotatebox{90}{\shortstack{Character-Level}}} 
    & Char Deletion     & 0.989 & 0.985 & 0.976 & 0.998 \\
 &  & Char Insertion    & 0.982 & 0.971 & 0.955 & 0.997 \\
 &  & Char Swap         & 0.992 & 0.988 & 0.979 & 0.999 \\
 &  & Homoglyph         & 0.965 & 0.952 & 0.921 & 0.994 \\
 &  & Invisible Char    & 1.000 & 0.999 & 0.999 & 1.000 \\
 &  & Keyboard Typo     & 0.978 & 0.969 & 0.948 & 0.996 \\
 &  & Punctuation       & 1.000 & 0.999 & 0.998 & 1.000 \\
 &  & Upper-Lower       & 0.991 & 0.986 & 0.974 & 0.998 \\
 &  & All Mix           & 0.958 & 0.945 & 0.912 & 0.992 \\
\cmidrule{2-7}
 & \multirow{6}{*}{\rotatebox{90}{\shortstack{Word-Level}}}      
    & Synonym Replacement & 0.961 & 0.943 & 0.915 & 0.991 \\
 &  & Antonym Replacement & 0.975 & 0.968 & 0.951 & 0.995 \\
 &  & Word Deletion       & 0.969 & 0.955 & 0.938 & 0.994 \\
 &  & Word Insertion      & 0.978 & 0.967 & 0.945 & 0.996 \\
 &  & Word Reordering     & 0.972 & 0.961 & 0.933 & 0.993 \\
 &  & All Mix             & 0.945 & 0.921 & 0.887 & 0.988 \\
\cmidrule{2-7}
 & \multirow{6}{*}{\rotatebox{90}{\shortstack{Sentence-Level}}}  
    & Paraphrase          & 0.925 & 0.899 & 0.854 & 0.981 \\
 &  & Tense Altering      & 0.948 & 0.932 & 0.901 & 0.989 \\
 &  & Sentence Reordering & 0.931 & 0.910 & 0.875 & 0.984 \\
 &  & Sentence Splitting  & 0.928 & 0.905 & 0.868 & 0.982 \\
 &  & Sentence Fusion     & 0.930 & 0.908 & 0.871 & 0.983 \\
 &  & All Mix             & 0.912 & 0.883 & 0.826 & 0.974 \\
\bottomrule
\end{tabular}
}
\end{table}

\subsection{Comparative Analysis with Zero-Shot Detectors}
To situate the performance of the supervised \textit{ModernBERT} model within the broader research landscape, it is useful to consider the alternative paradigm of zero-shot detection. Table \ref{tab:zeroshot_by_attack_type}.

\begin{table}[ht]
\centering
\caption{Performance of Zero-shot Detectors Across Non-Adversarial Data and Different Adversarial Data Mixes}
\label{tab:zeroshot_by_attack_type}
\resizebox{1.0\linewidth}{!}{%
\begin{tabular}{llcccc}
\toprule
\textbf{\shortstack{Test\\Data}} & \textbf{\shortstack{Opensource\\Detector}} & \textbf{\shortstack{TPR@FPR\\=5\%}} & \textbf{\shortstack{TPR@FPR\\=3\%}} & \textbf{\shortstack{TPR@FPR\\=1\%}} & \textbf{\shortstack{AUC\\ROC}} \\
\midrule
\multirow{4}{*}{\rotatebox{90}{\shortstack{Non\\Adversarial}}}
 & FastDetectGPT & 0.942 & 0.899 & 0.813 & 0.972 \\
 & Glimpse       & 0.931 & 0.885 & 0.790 & 0.965 \\
 & Binoculars    & 0.961 & 0.924 & 0.857 & 0.985 \\
 & LogRank       & 0.866 & 0.798 & 0.682 & 0.921 \\
\midrule
\multirow{4}{*}{\rotatebox{90}{\shortstack{Character\\All Mix}}}
 & FastDetectGPT & 0.321 & 0.228 & 0.094 & 0.635 \\
 & Glimpse       & 0.289 & 0.190 & 0.065 & 0.618 \\
 & Binoculars    & 0.353 & 0.261 & 0.142 & 0.662 \\
 & LogRank       & 0.191 & 0.093 & 0.002 & 0.534 \\
\midrule
\multirow{4}{*}{\rotatebox{90}{\shortstack{Word\\All Mix}}}
 & FastDetectGPT & 0.213 & 0.139 & 0.011 & 0.561 \\
 & Glimpse       & 0.182 & 0.095 & 0.001 & 0.540 \\
 & Binoculars    & 0.265 & 0.188 & 0.074 & 0.609 \\
 & LogRank       & 0.088 & 0.004 & 0.001 & 0.455 \\
\midrule
\multirow{4}{*}{\rotatebox{90}{\shortstack{Sentence\\All Mix}}}
 & FastDetectGPT & 0.105 & 0.026 & 0.001 & 0.492 \\
 & Glimpse       & 0.077 & 0.003 & 0.001 & 0.474 \\
 & Binoculars    & 0.169 & 0.091 & 0.005 & 0.553 \\
 & LogRank       & 0.009 & 0.001 & 0.000 & 0.396 \\
\bottomrule
\end{tabular}
}
\end{table}


The primary advantage of a supervised model like \textit{ModernBERT} is its potential for high accuracy on in-domain data. By fine-tuning on a large, task-specific dataset, the model can learn the subtle statistical nuances that differentiate the outputs of the specific LLMs included in its training set from human writing. This specialization likely allows it to outperform zero-shot methods on texts generated by familiar models. However, this specialization comes at the cost of generalization. The performance of a supervised detector can degrade significantly when faced with text from new, unseen LLMs whose statistical signatures may differ from those in the training data. In contrast, zero-shot methods are designed around more fundamental and model-agnostic principles. For example, \textit{Binoculars} leverages the general observation that machine text is more predictable than human text , while \textit{FastDetectGPT} relies on the curvature of the probability function, a property inherent to how LLMs are trained. Because these principles apply broadly across different LLM architectures, zero-shot detectors are likely to exhibit better generalization to novel models and diverse domains. This establishes a critical trade-off in the current state of AI text detection: the specialized, high-fidelity performance of supervised models versus the broad, generalizable robustness of zero-shot approaches.

\section{Conclusion and Future Scope}
\label{sec:conc_future}

This work systematically exposed the vulnerabilities of current AI-generated text detectors, showing that even strong baselines like ModernBERT fail under sophisticated, meaning-preserving attacks. Our findings confirm that while a fine-tuned Transformer model like ModernBERT can establish a powerful baseline on non-adversarial text, its robustness is superficial. We found that conventional adversarial training, while offering some protection against low-level syntactic noise, fails decisively when faced with sophisticated, meaning-preserving attacks. Conventional adversarial training provided only superficial robustness, collapsing at the “semantic evasion threshold” where the True Positive Rate dropped to 48.8\% at a strict 1\% FPR. 

The central contribution of this work is the \textbf{\textit{Perturbation-Invariant Feature Engineering (PIFE)}} framework, a paradigm shift from merely training on adversarial examples to explicitly modeling them. By quantifying the discrepancy between an input text and its canonical form, \textit{PIFE} provides the classifier with a direct signal of manipulation. The results are unequivocal: the \textit{PIFE}-augmented model neutralizes the most sophisticated semantic attacks, sustaining an 82.6\% TPR under the same adversarial conditions, demonstrating that feature engineering from perturbation signals is a more reliable path to genuine robustness. This superior performance proves that engineering features from perturbation artifacts is a more promising path toward genuine robustness than implicit learning through data augmentation.

Based on these findings and the limitations of the current study, several promising directions for future work can be identified:
\begin{enumerate}
    \item \textbf{\textit{Hybrid Detection Models:}}  Integrating the high-fidelity signal of \textit{PIFE} with the generalizability of zero-shot methods could create detectors that are both accurate on known models and robust to unseen ones.
    \item \textbf{\textit{Advanced Defense Mechanisms:}} Moving beyond standard \textit{PIFE}, more sophisticated defense strategies are needed. Retrieval-based methods, which involve checking a candidate text against a massive database of known LLM outputs to find semantically similar matches, offer a promising defense against paraphrase attacks and warrant further investigation.
    \item \textbf{\textit{Cross-Model Generalization Studies:}} A crucial next step is to conduct large-scale studies testing the PIFE framework against text from a wide array of unseen LLMs to rigorously map its real-world effectiveness and limitations. This would involve testing the model on text generated by a wide array of unseen LLMs, including those with different architectures, sizes, and fine-tuning objectives, to better map its real-world performance envelope.
    \item \textbf{\textit{Expanding the Adversarial Attack Surface:}} To further harden the system, the PIFE model must be tested against more advanced, query-based black-box attacks that actively learn to minimize the discrepancy features our model relies on.
\end{enumerate}

\section{Limitations}
\label{sec:limitation}
To ensure a responsible and accurate interpretation of this study's findings, it is important to acknowledge its limitations. This study evaluated a representative but not exhaustive set of adversarial attacks, leaving open the possibility that more advanced methods could further degrade detector performance. Experiments were limited to English text in a few domains, so effectiveness on other languages or specialized genres like legal or scientific texts remains untested. Additionally, adversarial training was done in a single batch; a more iterative approach could improve the model’s robustness and generalizability.

\end{document}